\documentclass{elsart}
\begin{document}

,%
\begin{frontmatter}
\title{Persistent Currents in Carbon Nanotubes}
\author{M. Szopa, M. Marga\'{n}ska\thanksref{email}}, \author{E. Zipper} 
\address{ul. Uniwersytecka 4, 40-007 Katowice, Poland}
\thanks[email]{corresponding author: magdalena@server.phys.us.edu.pl}

\begin{abstract}
Persistent currents driven by a static magnetic flux parallel to the
carbon nanotube axis are investigated. Owing to the hexagonal symmetry of
graphene the Fermi contour expected for a 2D-lattice reduces to two points.
However the electron or hole doping shifts the Fermi energy upwards or
downwards and as a result, the shape of the Fermi surface changes. Such a
hole doping leading to the Fermi level shift of (more or less) 1eV has been
recently observed experimentally. In this paper we show that the shift of
the Fermi energy changes dramatically the persistent currents and discuss
the electronic structure and possible currents for zigzag as well as
armchair nanotubes.
\end{abstract}
\end{frontmatter}

\section{Introduction}

There is a considerable research activity to explain physical properties
of single-wall and multi-wall carbon nanotubes. 
Their remarkable magnetic and electrical properties 
stem from the unusual electronic
structure of the graphene sheets - the quasi-2D material from which they are
made. The influence of
magnetic field on transport properties of carbon nanotubes has been explored,
both in theory \cite{ando,roc1,roc2} and in experiment \cite{fuji,lee,kim}.
In our paper we consider the influence of the magnetic field on persistent 
currents and structural properties of these materials induced by their 
topology.
The strong dependence of the band structure on the magnetic field
suggests large orbital magnetic response. Indeed, large magnetic
susceptibilities were calculated and experimentally observed \cite{bysz,lu}
for magnetic fields both perpendicular and parallel to the tube axis. These
results suggest the existence of persistent currents.\newline
In a nanotube the component of the momentum along the circumference of the
tube is quantized owing to the periodic boundary conditions. This raises the
possibility of inducing persistent currents along the circumference of the
nanotube by a static magnetic field applied parallel to the tube axis. They
arise due to Aharonov-Bohm effect. Such currents, but in toroidal nanotubes,
have been recently discussed theoretically \cite{lin,marg}.\newline
Persistent currents in nanotubes bear some resemblance with those in
metallic and semiconducting rings or cylinders. It is well known that
persistent currents in mesoscopic cylinders depend strongly on the
correlation of currents from different channels \cite{ibm} and on the shape
of Fermi surface \cite{steb}. The magnitude of the currents increases with
increasing curvature of the Fermi surface. The current is strongest for the
Fermi surface having large flat regions perpendicular to the direction in
which the magnetic field changes the wave vector $\mathbf{k}$.\newline
The shape of the Fermi contour expected for a 2D hexagonal lattice depends
strongly on the band filling. For an undoped carbon nanotube $E_{F}=0$ and
the Fermi surface reduces to two isolated Fermi points instead of a whole
contour. Electron or hole doping shifts the Fermi energy upwards or
downwards. As a result, the shape of the Fermi surface changes and this
influences the persistent currents. A substantial Fermi level shift of (more
or less) 1eV has been recently obtained \cite{krue} by the means of an
electrochemical gating. \newline
In the present paper we study peristent currents induced by a static
magnetic field parallel to the axis of a single wall carbon nanotube. We
investigate the magnitude and shape of the current as as well as their
dependence on the hole doping. In particular we show that for zigzag
nanotubes, for certain values of the hole doping, we can obtain a large
enhancement of the persistent current. Finally, we briefly discuss the
possibility of creating self-sustaining currents in multiwall nanotubes.

\section{The structure of a nanotube}

The nanotube is a rolled up strip of a graphene sheet. The rolling vector is 
$\mathbf{L}_{t}=m_{1}\mathbf{T}_{1}+m_{2}\mathbf{T}_{2}$ (the $t$ index
stands for transverse direction), where $\mathbf{T}_{1}$ and $\mathbf{T}_{2}$
form the standard basis vectors of the graphene lattice \cite{nano}. The
vector of the length of the nanotube is $\mathbf{L}_{l}=p_{1}\mathbf{T}%
_{1}+p_{2}\mathbf{T}_{2}$ (the $l$ index stands for longitudinal direction).
The four integers $(m_{1},m_{2})\times (p_{1},p_{2})$ therefore define the
geometry of the tube. The number of lattice nodes of the nanotube is $%
N=m_{1}p_{2}-m_{2}p_{1}.$ Note that $\mathbf{L}_{t}$ and $\mathbf{L}_{l}$ do
not have to be orthogonal (the nanotube is then twisted \cite{nano}) but for
simplicity we always take in our paper the twist angle to be 0, i.e. $%
m_{1},m_{2},p_{1},p_{2}$ such that they are orthogonal.\newline
The wave vector $\mathbf{k}=(k_{x},k_{y})$ of an electron on a surface of a
2D cylindrical tube obeys two boundary conditions. The first, periodic along
the circumference of the nanotube, is 
\begin{equation}
\mathbf{k}\cdot \mathbf{L}_{t}=k_{t}|\mathbf{L}_{t}|=2\pi l_{t}^{\prime },
\label{klt}
\end{equation}%
where, in the absence of magnetic field, $l_{t}^{\prime }$ is an integer.
The second boundary condition, which results in vanishing of the sinusoidal
wave function at the ends of the nanotube, is 
\begin{equation}
\mathbf{k}\cdot \mathbf{L}_{l}=k_{l}|\mathbf{L}_{l}|=\pi l_{l},  \label{kll}
\end{equation}%
where $l_{l}$ is an arbitrary integer. The first Brillouin zone of the whole
graphene sheet is a hexagone whose vertices are $(\pm \frac{2\pi }{3\sqrt{3}}%
,\pm \frac{2\pi }{3})$ and $(0,\pm \frac{4\pi }{3})$. The wave functions
with $l_{l}$ and  $-l_{l}$ are linearly dependent and therefore the first
Brillouin zone of the nanotube reduces to a half of the original hexagone.
In our choice $l_{l}$ takes only positive values. 

The energy of the $\mathbf{k}$-th state of the hexagonal lattice, in the
tight binding approximation, is well known \cite{nano} to be 
\begin{equation}
E_{\mathbf{k}}=\pm \gamma \sqrt{1+4\cos ^{2}\frac{\sqrt{3}}{2}k_{x}+4\cos 
\frac{\sqrt{3}}{2}k_{x}\frac{3}{2}k_{y}},  \label{ek}
\end{equation}%
where $\gamma $ is the resonance integral for the nearest neighbour
interaction (we take $\gamma =3.033eV,$ typical for the carbon hexagonal
lattice \cite{lin}). The energy may change from $-3\gamma $ to $3\gamma $.
At the half-filling, when half of the possible states are occupied, the
Fermi level is at $E_{F}=0$ and the Fermi surface reduces to isolated points
-- vertices of the original hexagone. Only two of them, e.g. $K_{1}=(-\frac{%
2\pi }{3\sqrt{3}},\frac{2\pi }{3})$ and $K_{2}=(\frac{2\pi }{3\sqrt{3}},%
\frac{2\pi }{3})$, are linearly independent and therefore define the whole
Fermi surface of the nanotube at $E_{F}=0$. When less than a half of the
states are occupied (the nanotube is hole-doped), the Fermi level lowers and
becomes a set of curves (drawn as cotinuous bold lines in Fig. 2--4a, and
Fig. 6a). At $E_{F}=-\gamma $ the Fermi surface is a half-hexagone whose
vertices are at the centre of edges of the first Brillouin zone (cf. Fig.
3a). In this case the number of occupied states $N_{e}$ is equal to 3/4 of
the half-filling, i.e. $N_{e}=\frac{3}{4}N$.

After applying the magnetic field along the axis of the nanotube, all the
wavefunctions acquire the Aharonov-Bohm phase factor $e^{i2\pi \phi /\phi
_{0}}$, where $\phi $ is the magnetic flux through the nanotube and $\phi
_{0}=\frac{hc}{e}$ is the flux quantum. This phase factor is equivalent to a
shift in the momentum quantum number $l_{t}^{\prime }$ 
\begin{equation}
l_t^{\prime }=l_t+\frac \phi {\phi _0},  \label{ltprim}
\end{equation}
where $l_t$ is an arbitrary integer. The magnetic field induces persistent
current along the transverse direction of the nanotube 
\begin{equation}
I_{\mathbf{k}}(\phi )=-c\frac{\partial E_{\mathbf{k}}}{\partial \phi }.
\end{equation}
Summing over all momentum states within the first Brillouin zone and below
the Fermi level, we obtain the overall current, induced by the magnetic
flux. 
\begin{equation}
I(\phi)=\sum_{\mathbf{k}_{occupied}}I_{\mathbf{k}}(\phi).
\end{equation}
When all the shifted states remain within the Brillouin zone and below the
Fermi level, the current is diamagnetic. If, due to the shift induced by $%
\phi $, some states leave the Brillouin zone or cross the Fermi level, and
others enter it or cross the Fermi level in the opposite direction,  we
witness a paramagnetic jump of the current \cite{cheu}. This total current
depends also on the temperature which suppresses its amplitude \cite%
{steb,cheu}, but for simplicity we assume here that $T=0$.

\section{Persistent currents in zigzag nanotubes}

We have performed numerical calculations for nanotubes of various
geometries. First we discuss the case of a $(14,0)\times(-100,200)$
nanotube. The Brillouin zone, at zero magnetic field, for this nanotube is
shown in Fig. 1a. It is equivalent to the upper (or lower) half of the
original Brillouin zone of the graphene sheet \cite{nano}.

In the half filling case ($E_{F}=0$) there is one conduction electron per
each lattice node i.e. $N_{e}=N$. (in our case $N=2800$). This nanotube is
not metallic and there are no allowed momentum states in any of the Fermi
points (the condition for metallicity of the tube is $(m_1-m_2)|_{mod\;3}=0$%
). The persistent current in units of $I_0=\frac{\gamma c}{\phi_0}=1.29\cdot
10^{-4}$ A, induced by the magnetic field is plotted in Fig. 1b. The current
is periodic with period $\phi_0$ and diamagnetic at small $\phi$.\newline

As the ref.\cite{krue} shows, the Fermi energy can be lowered by binding
some of the electrons (hole doping) and then $N_{e}<N$. In the following we
assume that the doping lowers the number of electrons to a constant $N_{e}$,
which does not depend on $\phi $. Therefore, as the distribution of the
momentum states in the Brillouin zone changes with $\phi $, the value of $%
E_{F}$ must be adjusted to keep $N_{e}$ independent of $\phi $. This is not
the case for $E_F=0$, where the number of states in the first Brillouin zone
is constant ($N_e=N$) and does not depend on $\phi$.

The Fig. 2a shows the reciprocal lattice and the Fermi contour (bold line)
for $N_{e}=2744$ at $\phi =0$. It corresponds to the Fermi energy shift of
approximately $-1eV$, (or $\sim -\gamma /3$) which has been obtained
experimentally \cite{krue}. The change of Fermi energy with $\phi /\phi _{0}$
is shown in the Fig. 2b. We see that the oscillations are around $-\gamma /3$
and have quite a substantial amplitude of approximately 0.07$\gamma$. The
induced persistent current Fig. 2c has now four times higher amplitude and a
different shape.

The most favourable situation for the enhancement of the current is when the
Fermi level is at the value of $-\gamma $. It corresponds to the number of
electrons $N_{e}=\frac{3}{4}N$. The Fermi surface then becomes the smaller
half-hexagon in the first Brillouin zone (cf Fig. 3a), with the momentum
lines parallel to its vertical edges. The oscillations of the Fermi level
(shown on Fig. 3b) are three orders of magnitude smaller compared to the
previous case of $N_{e}=2744$. At $\phi=0$, a whole momentum line crosses
the Fermi surface with the change of $\phi$, which results in a large
paramagnetic jump of the current (Fig. 3c).

The current's amplitude is 25-30 times higher compared with the current at
the half-filling. This is the largest persistent current possible to obtain
in this nanotube. By further shifting the Fermi level downwards, the Fermi
surface becomes curved and again the amplitude of the current gradually
decreases.\newline

The amplitude of persistent currents, at the most favourable value of $3/4$
of half-filling, depends on the length and width of the tube. It is
proportional to its length and decreases with its width. The reason for this
is that the longer the tube is, the more dense are the states in the
longitudinal direction, therefore, when a line of them crosses the Fermi
surface, the effect is proportionally greater. On the other hand, the wider
the tube is, the smaller are individual currents: 
\begin{equation}  \label{indiv}
I_{\mathbf{k}}= -c\frac{\partial{E_{\mathbf{k}}}}{\partial{\phi}}= -c\frac{%
\partial E_{\mathbf{k}}}{\partial{k_t}}\frac{2\pi} {\phi_0|\mathbf{L}_t|}.
\end{equation}
Let us notice also what happens if the $m_{1}$ parameter of the nanotube
(the one which determines its diameter) changes. If $m_{1}$ is even, as in
the previously discussed case, there is an even number of $k_{t}=const$
state lines inside the Fermi contour for $N_{e}=\frac{3}{4}N$ (as in Fig.
3a) and the paramagnetic jump of the current (Fig. 3c) occurs at integer
values of $\phi /\phi _{0}$. If $m_{1}$ is odd, the number of $k_{t}=const$
state lines inside the Fermi contour is odd and the paramagnetic jump of the
current occurs at half-integer values of $\phi /\phi _{0}$. We investigated
such a situation and the result for a $(15,0)\times(-100,200)$ nanotube is
illustrated in Fig. 4. Again a strongly enhanced current is obtained, but
this time it is diamagnetic at small $\phi$. This behaviour of the current
bears a strong resemblance to what happens in a one-dimensional metallic
ring:  the currents like the one in Fig. 3c (``even'' nanotube) occur if the
number of electrons in the ring is even and the case from the Fig. 4c
(``odd'' nanotube) takes place for an odd number of electrons \cite{cheu}.
This analogy arises from the equivalent positions of the Fermi level with
respect to the last occupied state (line of states) in these two systems.%
\newline

\section{Persistent currents in armchair nanotubes}

At the half filling, the principal difference between nanotubes of different
chiralities is whether they are metallic or semi-conducting. We know that if
the nanotube parameters obey the relation $(m_1-m_2)|_{mod\;3}=0$, the tube
is metallic, otherwise it is semiconducting. This criterion follows from the
specific structure of the Brillouin zone at half-filling, and is not valid
for $E_F\neq0$. At lower, or higher, band filling all the nanotubes are more
likely to be metallic because when the Fermi points become lines, the
probability that some states lie on them increases.

The important feature is the shape of the tube's momentum spectrum. The
chirality of the tube varies from $(m,0)$ for zigzag nanotubes (where the
momentum lines are parallel to the sides of the smaller, $E_{F}=-\gamma$
hexagon) to $(m,m)$ for armchair nanotubes (where they are parallel to one
of the sides of the Brillouin zone). We now take the case of a $(9,9)\times
(-150,150)$ nanotube which is shown in Fig. 5. Its length and width are
similar to the $(14,0)\times (-100,200)$ tube and the number of electrons in
the half filling case is $N_{e}=N=2700$. The momentum lines are parallel to
the sides of the Brillouin zone and the resulting current (Fig. 5b) has
amplitude of the order of $2I_0$.\newline
The current in this case is much smaller compared to the similar
distribution of $k$-states within the allowed region in the Fig. 3a. The
reason for the attenuation of the current in this case is that the
individual currents (\ref{indiv}) corresponding to the states at the edge of
the Brillouin zone are much smaller than the currents carried by the states
at the edge of the Fermi surface. This can be seen by observing the
equienergy lines at these edges (Fig. 3a and 5a).\newline

Decreasing the number of electrons or lowering the Fermi level does not
enhance the persistent currents in this case. The reason for that is that
the momentum lines $k_t=const$ are not parallel to any fragment of the Fermi
surface. An example of such a situation is shown in Fig. 6, where where $N_e=%
\frac{3}{4}N$. Both the attenuation and the oscillations of the current, as
compared to the half filling case, are caused by the lack of correlation
between the states which now travel across the Fermi surface separately.%
\newline
In real systems, the temperature $T>0$ would smooth down the oscillations
shown in Fig. 6c and the resulting current would be very small.

\section{Conclusions}

The presence of quantized, delocalized electron states in carbon nanotubes
should result in persistent currents. In carbon nanotubes, which are
typically almost free of defects, these currents should be substantial and
can change with doping. We have investigated the changes of the shape of the
Fermi surface with the hole doping and discussed the respective changes of
persistent currents in zigzag as well as in armchair nanotubes. \newline
The currents are calculated under the assumption that the number of
electrons is constant in the nanotube. They can be paramagnetic or
diamagnetic depending on the position of states in the first Brillouin zone
and the shape of the Fermi surface. We have shown that the largest currents
can be obtained if the Fermi energy is lowered from $E_{F}=0$ (half filling
case) to $E_{F}\cong -1\gamma $ ($N_{e}=\frac{3}{4}N$) for zigzag nanotubes.
The Fermi surface changes then from two separate points to a flat contour
and the induced persistent currents increase their amplitude 25-35 times.
For chiral nanotubes such that $m_2<<m_1$ the enhancement of the persistent
current at $E_F=-\gamma$ can also be large. \newline
In armchair nanotubes the situation is different, namely with hole doping
the amplitude the peristent current decreases because the individual
currents become less correlated.

The large amplitude of the obtained currents raises the possibility of
obtaining self sustaining currents which produce the magnetic flux capable
to maintain themselves even in absence of external magnetic field \cite{wohl}%
. However, spontaneous flux in single wall nanotubes will be very small -- $%
\phi/\phi_0\simeq10^{-3}$ only. To discuss the problem of self-sustaining
currents multiwall nanotubes are more suitable. This problem will be
discussed in a forthcoming paper. Below we will only briefly mention various
possible cases.\newline
In the case of multi-wall nanotubes, where many tubes are arranged in a
coaxial fashion, the currents produced by all tube layers should superpose.
However, the electrical properties of individual tubes have been shown to
vary strongly from tube to tube \cite{scho}. First, let us assume that all
the tubes have the same Fermi energy $E_F=-\gamma$. The most favourable case
is when all (or a majority of) the tubes are of the zigzag type discussed in
the section 3.\newline
If all tubes have the same parity (even or odd) the amplitude of the total
current is greatest -- it can be 10-50 times larger than in a single-wall
nanotube. This current is paramagnetic for even and diamagnetic for odd $%
m_{1}$ for small values of $\phi /\phi _{0}$. In such situation there is a
large probability of obtaining self-sustaining currents.\newline
When tubes belonging to the multiwall nanotube have different chiralities,
the total current will be smaller due to the cancellation between currents
from different tubes. However, the total current will not average to zero,
and the net paramagnetic current with period halving will be obtained. The
least favourable case is when none of the tubes are of the zigzag type. The
cancellation is then substantial and we obtain only a very small net current.%
\newline
It seems that these novel and unusual properties of carbon nanotubes may
have promising applications in areas such as the magnetoelectronics, e.g. in
magnetic sensors.\newline

\noindent{Work supported by KBN grant Nr 5PO3503320.}

\newpage
\begin{figure}
\vspace{9cm} 
\begin{picture}(0,0)
\put(10,180){\small{a)}}
\put(345,110){\footnotesize{$k_t$}}
\put(27,254){\footnotesize{$k_l$}}
\put(80,60){\small{b)}}
\put(280,40){\scriptsize{$\phi/\phi_0$}}
\put(110,85){\scriptsize{$I/I_0$}}
\end{picture} 
\includegraphics{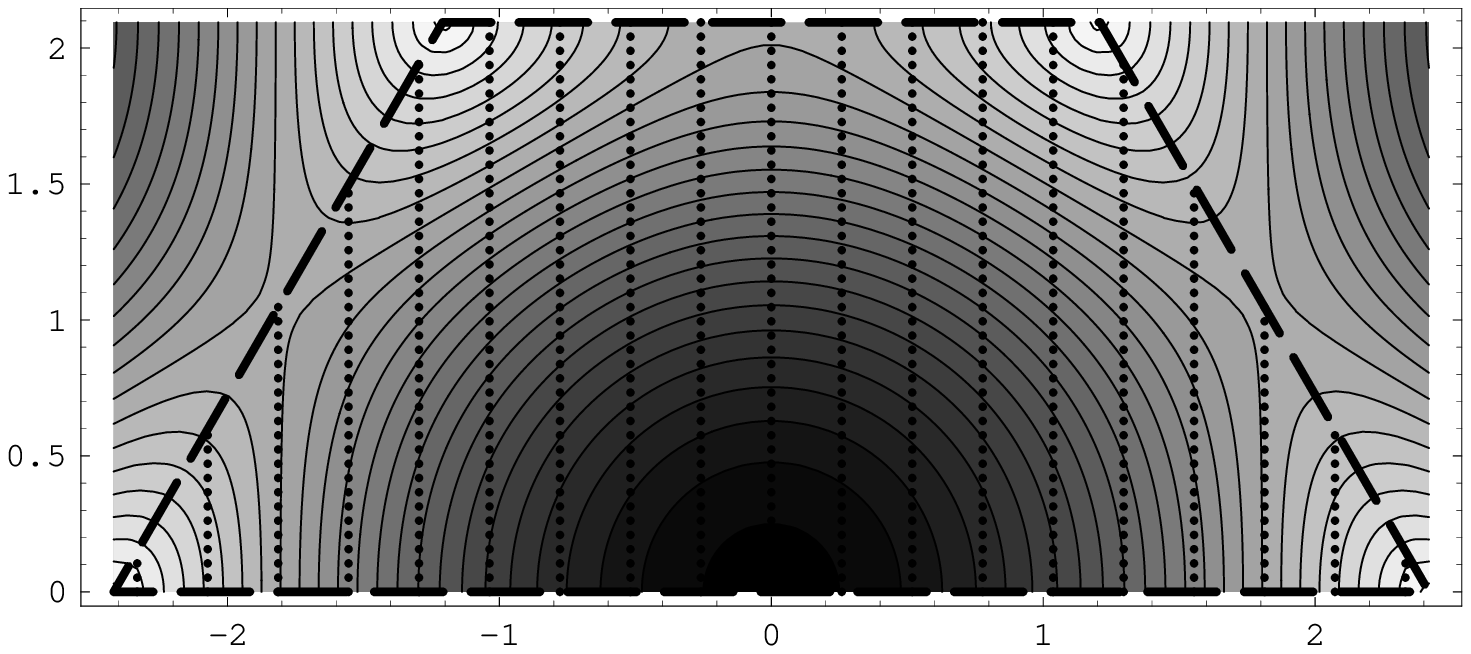} \includegraphics{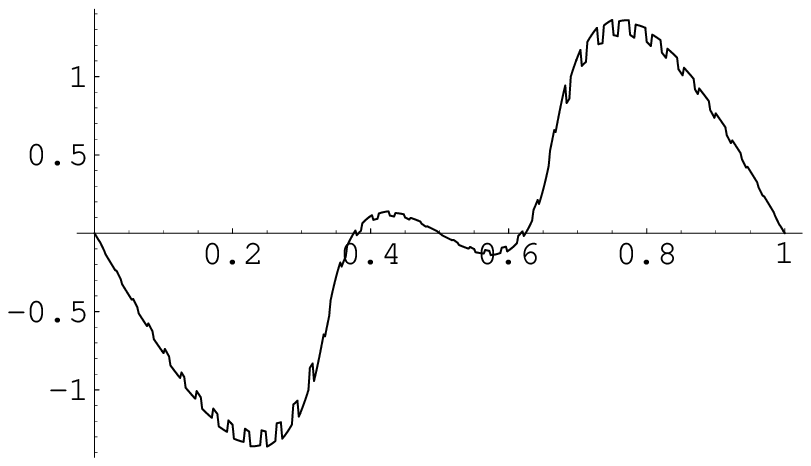}
\caption{a) The Brillouin zone of a $(14,0)\times(-100,200)$ nanotube at $%
\phi=0$ and $N_e=2800$ (half-filling). The dotted lines represent different $%
k_t=const$ momentum lines. The shading in the background is the contour plot
of the $E(\mathbf{k})$ relation \ref{ek}. The dashed line is the edge of the
Brillouin zone. b) The persistent current $I(\phi/\phi_0)/I_0$.}
\end{figure}
\newpage
\begin{figure}
\vspace{8.7cm} 
\begin{picture}(0,0)
\put(30,190){\small{a)}}
\put(330,110){\footnotesize{$k_t$}}
\put(55,225){\footnotesize{$k_l$}}
\put(270,70){\small{c)}}
\put(380,48){\scriptsize{$\phi/\phi_0$}}
\put(210,90){\scriptsize{$I/I_0$}}
\put(100,40){\small{b)}}
\put(180,73){\scriptsize{$\phi/\phi_0$}}
\put(30,90){\scriptsize{$E_F$}}
\end{picture} 
\includegraphics{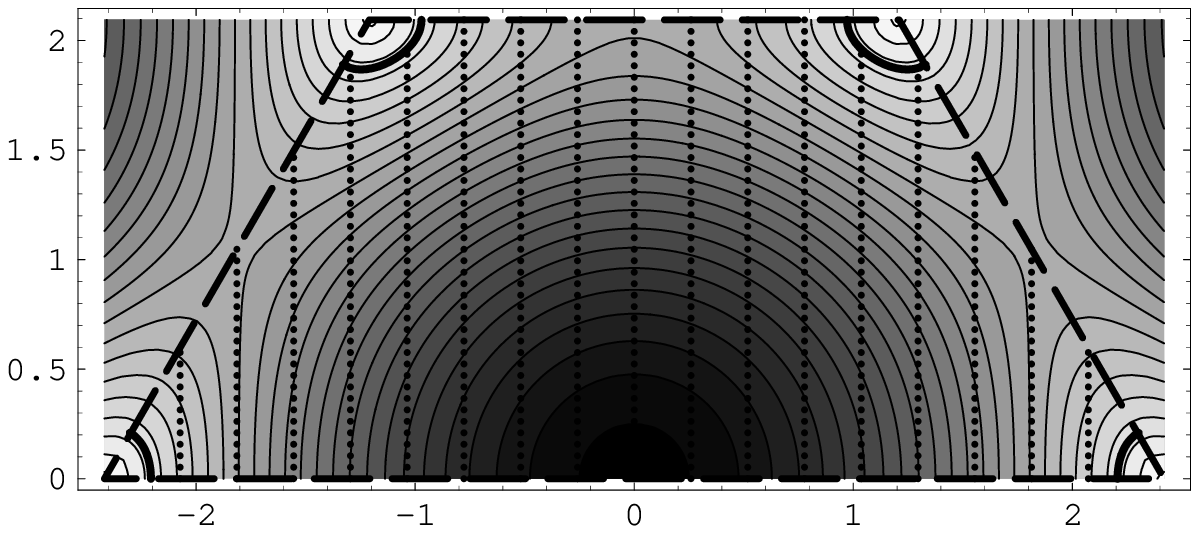} \includegraphics{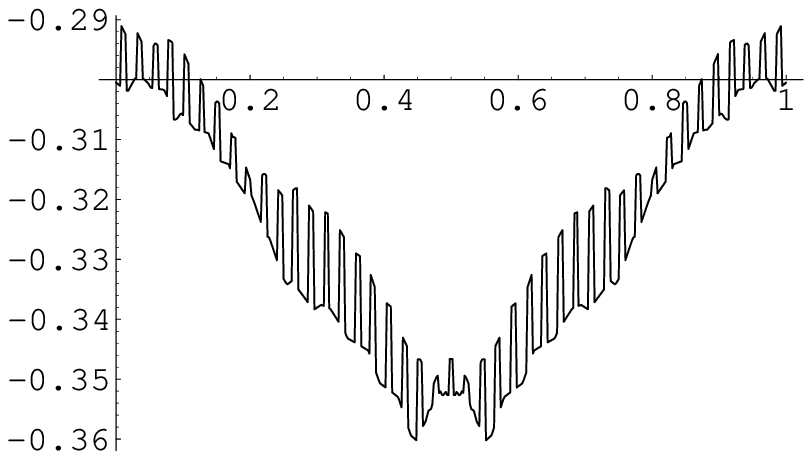} \includegraphics{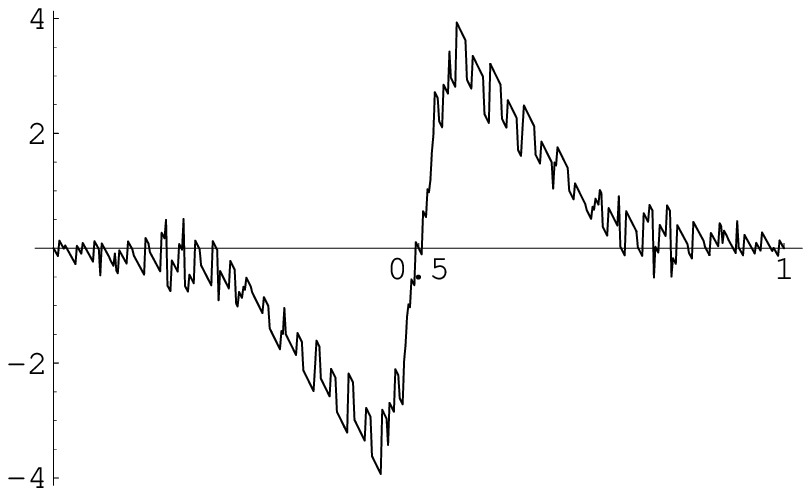}
\caption{a) The Brillouin zone and Fermi surface, b) the $E_{F}(\phi/\phi_{0})$
plot and c) the $I(\phi/\phi _{0})$ for a $(14,0)\times (-100,200)$ nanotube
at $N_{e}=2744$ i.e. the Fermi energy lowered by about $\gamma /3$. The
continuous bold contour on a) is the Fermi surface at $\phi=0$.}
\end{figure}
\newpage
\begin{figure}
\vspace{8.5cm} 
\begin{picture}(0,0)
\put(10,180){\small{a)}}
\put(355,103){\footnotesize{$k_t$}}
\put(20,255){\footnotesize{$k_l$}}
\put(320,70){\small{c)}}
\put(375,48){\scriptsize{$\phi/\phi_0$}}
\put(205,90){\scriptsize{$I/I_0$}}
\put(100,55){\small{b)}}
\put(180,48){\scriptsize{$\phi/\phi_0$}}
\put(45,85){\scriptsize{$E_F$}}
\end{picture} 
\includegraphics{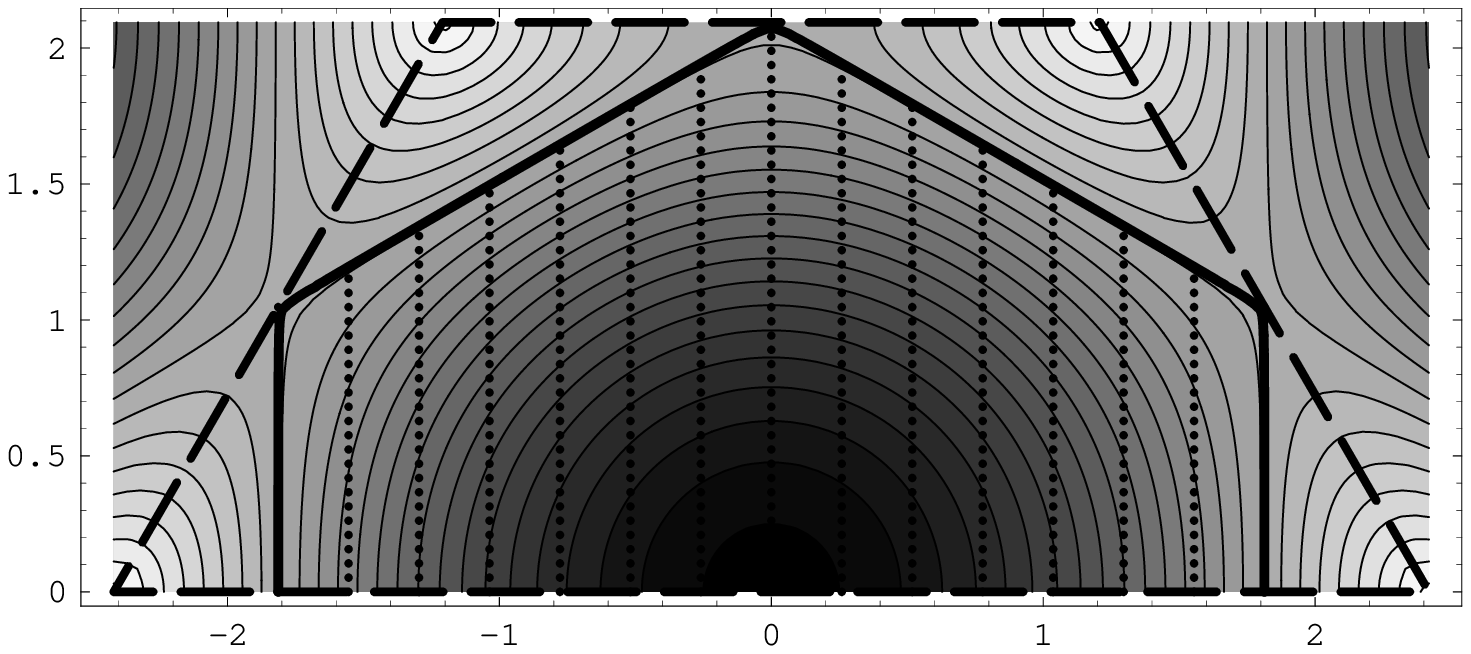} \includegraphics{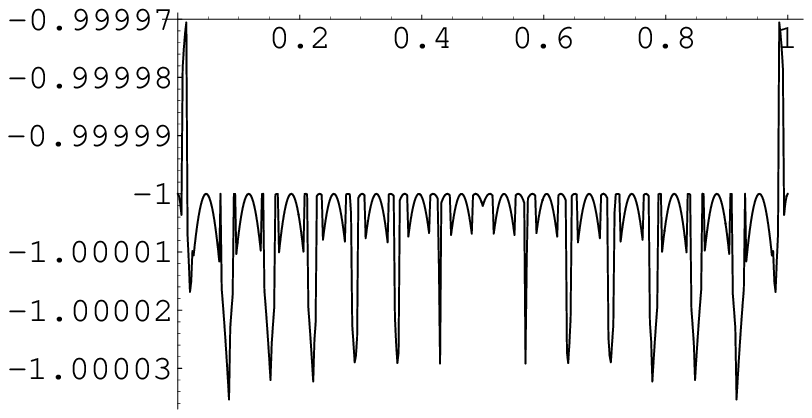} \includegraphics{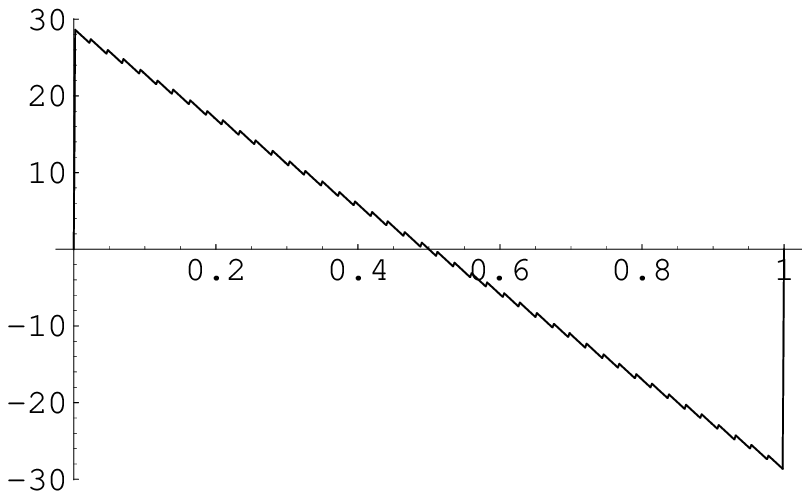}
\caption{a) The plot represents the Brillouin zone and the Fermi surface (bold
contour) of a $(14,0)\times (-100,200)$ nanotube at $N_{e}=2100=\frac{3}{4}N$%
; b) is the plot of Fermi energy vs magnetic flux; c) is the resulting
persistent current vs magnetic flux.}
\end{figure}
\newpage
\begin{figure}
\vspace{9.2cm} 
\begin{picture}(0,0)
\put(10,200){\small{a)}}
\put(360,100){\footnotesize{$k_t$}}
\put(30,245){\footnotesize{$k_l$}}
\put(260,65){\small{c)}}
\put(380,40){\scriptsize{$\phi/\phi_0$}}
\put(210,85){\scriptsize{$I/I_0$}}
\put(100,10){\small{b)}}
\put(180,43){\scriptsize{$\phi/\phi_0$}}
\put(30,85){\scriptsize{$E_F$}}
\end{picture} 
\includegraphics{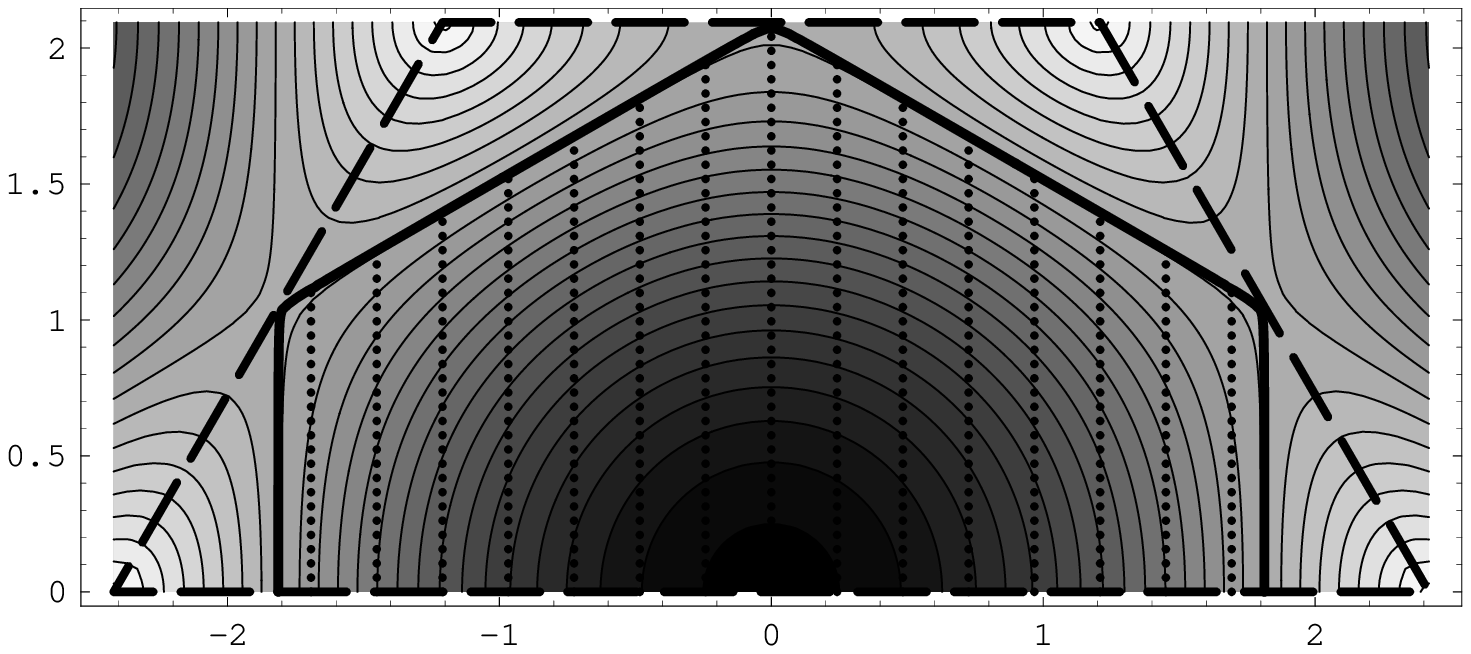} \includegraphics{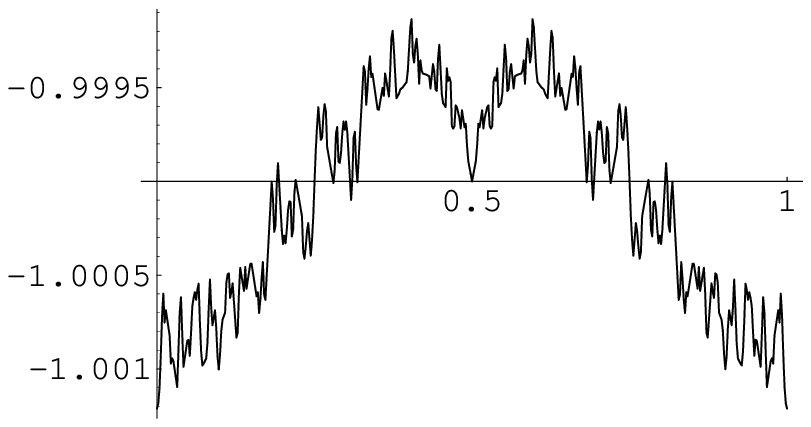} \includegraphics{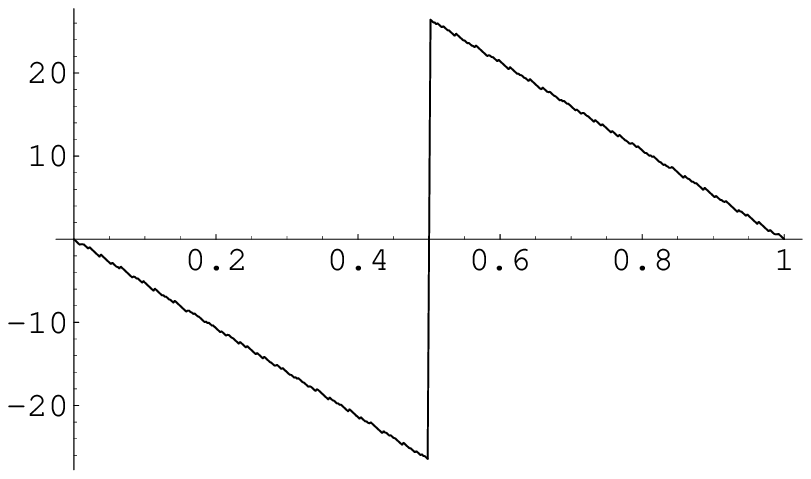}
\caption{The plots are all for a nanotube at $N_e=3/4 N_0$. a) The Brillouin
zone and the Fermi surface (bold line) for a $(15,0)\times(-100,200)$
nanotube at $\phi=0$ -- note that there is no momentum line at the Fermi
level. b) The change of Fermi energy with $\phi$. c) The resulting
persistent current - note that the jump occurs at $\phi/\phi_0=1/2$.}
\end{figure}
\newpage
\begin{figure}
\vspace{4.5cm} 
\begin{picture}(0,0)
\put(-10,105){\small{a)}}
\put(240,0){\footnotesize{$k_t$}}
\put(0,125){\footnotesize{$k_l$}}
\put(320,100){\small{b)}}
\put(375,70){\scriptsize{$\phi/\phi_0$}}
\put(250,110){\scriptsize{$I/I_0$}}
\end{picture}
\includegraphics{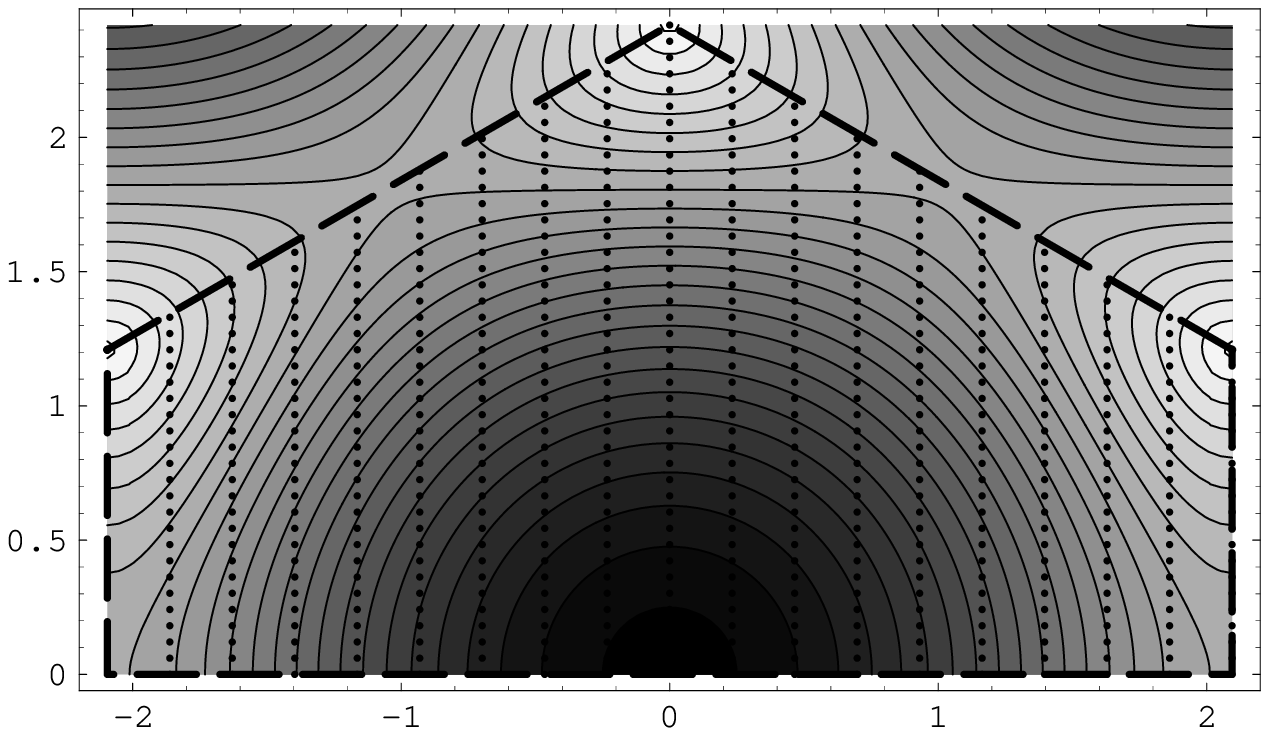} \includegraphics{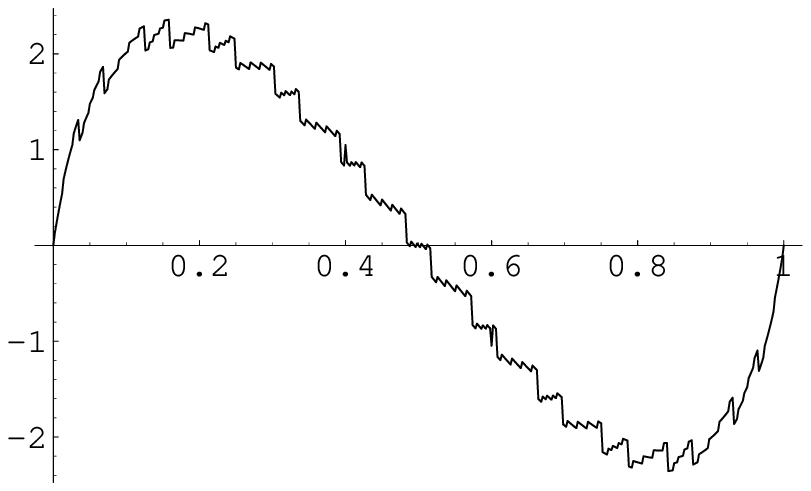}
\caption{a) The Brillouin zone and momentum states for a $ (9,9)
\times(-150,150)$ nanotube at half-filling, i.e. $N_e=2700$. b)  Persistent
currents vs magnetic flux, everything at constant number of electrons.}
\end{figure}
\newpage
\begin{figure}
\vspace{8.5cm} 
\begin{picture}(0,0)
\put(55,230){\small{a)}}
\put(325,120){\footnotesize{$k_t$}}
\put(70,260){\footnotesize{$k_l$}}
\put(280,80){\small{c)}}
\put(380,40){\scriptsize{$\phi/\phi_0$}}
\put(225,95){\scriptsize{$I/I_0$}}
\put(100,70){\small{b)}}
\put(185,30){\scriptsize{$\phi/\phi_0$}}
\put(30,97){\scriptsize{$E_F$}}
\end{picture} 
\includegraphics{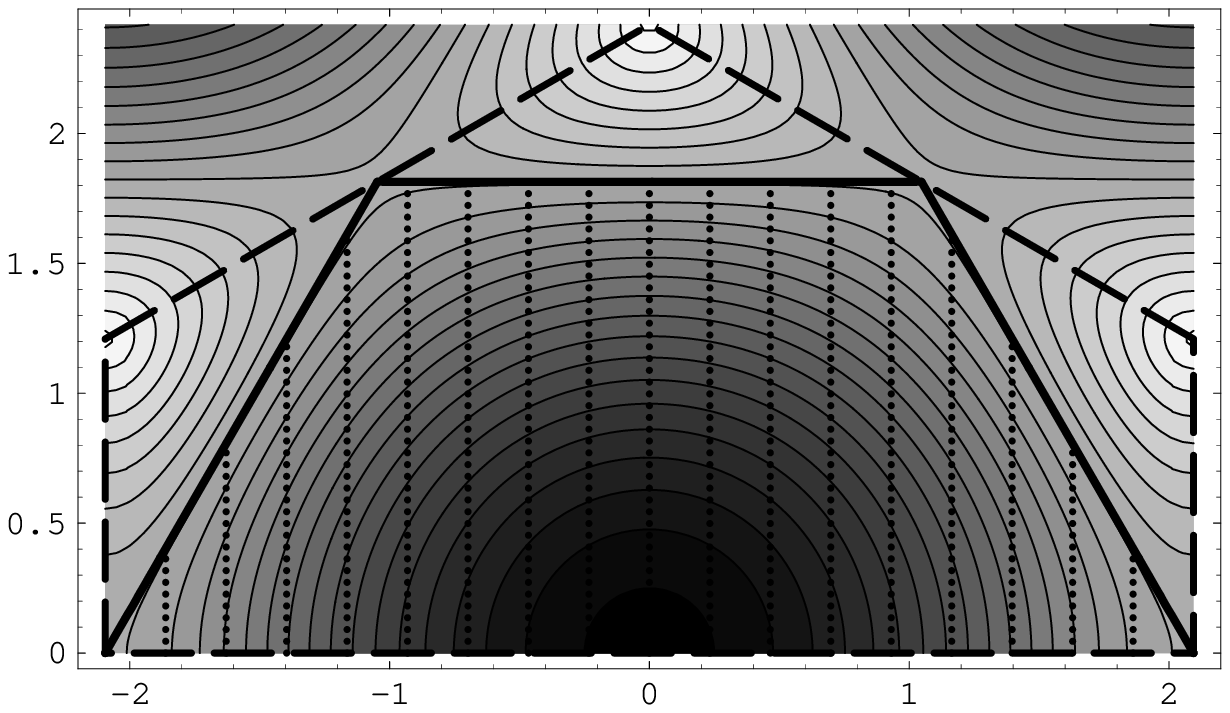} \includegraphics{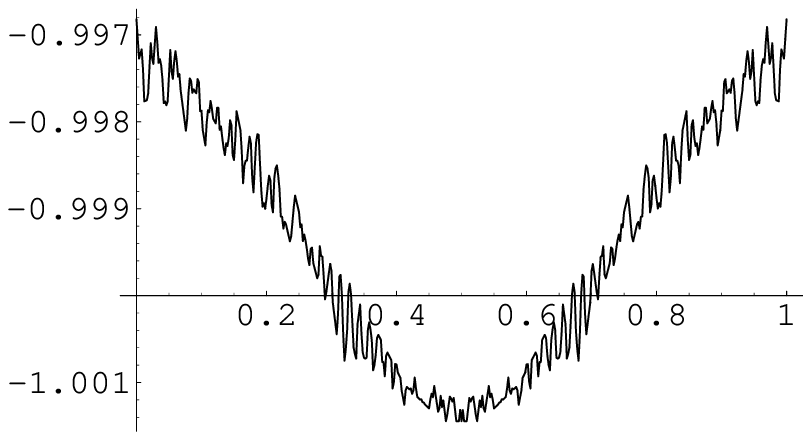} \includegraphics{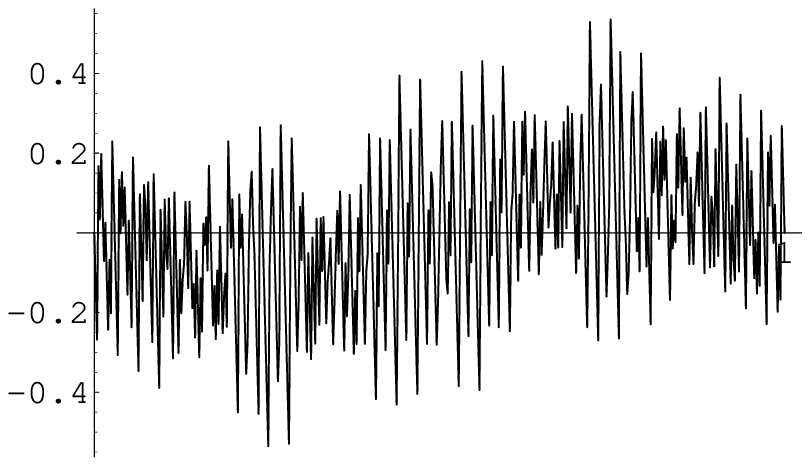}
\caption{a) Brillouin zone and the Fermi surface of a $(9,9)\times (-150,150)$
nanotube at $N_{e}=\frac{3}{4}N$. b) $E_{F}(\phi/\phi _{0})$ plot, c) $%
I(\phi /\phi _{0})$.}
\end{figure}


\begin{thebibliography}{99}
\bibitem{ando} T. Ando, Semicond. Sci. Technol. (2000) R13

\bibitem{roc1} S. Roche, G. Dresselhaus, M.S. Dresselhaus, R. Saito, Phys. Rev. B \textbf{62} (2000) 16092

\bibitem{roc2} S. Roche, R. Saito, Phys. Rev. Lett \textbf{87} (2001) 246803

\bibitem{fuji} A. Fujiwara {\em et al.}, Phys. Rev. B \textbf{60} (1999) 13492

\bibitem{lee} J.O. Lee {\em et al.}, Phys. Rev. B \textbf{61} (2000) R16362

\bibitem{kim} N. Kim {\em et al.}, J. Phys. Soc. Jpn \textbf{70} (2001) 789 

\bibitem{bysz} P. Byszewski and M. Baran, Europh. Letters \textbf{31} (1995)
363

\bibitem{lu} J.P.Lu, Phys. Rev. Lett. \textbf{74} (1995) 1123

\bibitem{lin} M. F. Lin, D. S. Chou, Phys. Rev B, \textbf{57} (1998) 6731

\bibitem{marg} M. Marganska, M. Szopa, Acta Phys. Pol. \textbf{32} (2001) 427

\bibitem{ibm} H. Cheung, Y. Gefen, E.K. Riedel, IBM J. Res. Develop. \textbf{%
32} (1988) 359

\bibitem{steb} M. Stebelski, M. Szopa, E. Zipper, Z.Phys B, \textbf{103}
(1997) 79

\bibitem{krue} M. Kr\"{u}ger, M.R. Buitelaar, T. Nussbaumer, C. Sch\"{o}%
nenberger, Appl. Phys. Lett. 78 (2001) 1291

\bibitem{nano} R. Saito, G. Dresselhaus, M.S. Dresselhaus {\em``Physical Properties of Carbon Nanotubes''}, Imperial College Press, London 1998



\bibitem{cheu} H. Cheung, Y. Gefen, E.K. Riedel, W.-H. Shih, Phys Rev B, 
\textbf{37} (1988) 6050

\bibitem{wohl} D. Wohlleben, M. Esser, P. Freche, E. Zipper, M. Szopa, Phys
Rev Lett, \textbf{66} (1991) 3191

\bibitem{scho} C. Sch\"{o}nenberger, L. Forr\'{o}, Physics World, June 2000,
p.37
\end{thebibliography}
\end{document}